\documentclass[aps,prb,amsmath,amssymb,twocolumn,groupedaddress,longbibliography,floats,final,postprint,superscriptaddress]{revtex4-1}
\usepackage{amsmath,amssymb,amsthm,enumitem}
\usepackage{relsize}
\usepackage[citecolor=blue,linkcolor=blue,colorlinks]{hyperref}
\usepackage{tikz}
\usetikzlibrary{arrows}
\usepackage{pgf}
\usepackage{pgfcore}
\usepackage{subfigure}
\usepackage{multirow}
\usepackage{booktabs}
\usepackage{array}
\usepackage{algorithm}
\usepackage{algorithmic}
\usepackage{threeparttable}
\usepackage[normalem]{ulem}

\usepackage{enumitem}
\setenumerate{fullwidth,itemindent=\parindent,%
listparindent=\parindent,itemsep=0ex,partopsep=0pt,parsep=0ex}

\usepackage{txfonts}

\usepackage{xcolor}
\definecolor{ForestGreen}{HTML}{228b22}
\definecolor{Red}{HTML}{DC143C}
\definecolor{Orange}{HTML}{CC6600}
\definecolor{Purple}{HTML}{6B00B3}

\def\mz{\mathcal{Z}}
\newcommand{\mg}{\mathcal{G}}

\newcommand{\mru}{R_\uparrow}
\newcommand{\mrd}{R_\downarrow}
\newcommand{\mi}{\mathtt{I}}
\newcommand{\mm}{\mathtt{M}}

\begin{document}

\title{Worm-algorithm-type Simulation of Quantum Transverse-Field Ising Model}

\author{Chun-Jiong Huang}
\author{Longxiang Liu}
\affiliation{Shanghai Branch, National Laboratory for Physical Sciences at Microscale
and Department of Modern Physics, University of Science and Technology
of China, Shanghai, 201315, China}
\affiliation{CAS Center for Excellence and Synergetic Innovation Center in Quantum
Information and Quantum Physics, University of Science and Technology
of China, Hefei, Anhui 230026, China}
\affiliation{CAS-Alibaba Quantum Computing Laboratory, Shanghai, 201315, China}

\author{Yi Jiang}
\email[]{jiangyi@ustc.edu.cn}
\affiliation{Department of Modern Physics,
University of Science and Technology of China, Hefei, Anhui 230026, China}

\author{Youjin Deng}
\email[]{yjdeng@ustc.edu.cn}
\affiliation{Shanghai Branch, National Laboratory for Physical Sciences at Microscale
and Department of Modern Physics, University of Science and Technology
of China, Shanghai, 201315, China}
\affiliation{CAS Center for Excellence and Synergetic Innovation Center in Quantum
Information and Quantum Physics, University of Science and Technology
of China, Hefei, Anhui 230026, China}
\affiliation{CAS-Alibaba Quantum Computing Laboratory, Shanghai, 201315, China}

\date{\today}

\begin{abstract}
We apply a  worm algorithm to simulate the quantum transverse-field Ising model in a
path-integral representation of which the expansion basis is taken as the spin
component along the external-field direction. In such a representation,  a
configuration can be regarded as a set of non-intersecting loops constructed by
``kinks" for pairwise interactions and spin-down (or -up) imaginary-time segments.
The wrapping probability for spin-down loops, a dimensionless quantity
characterizing the loop topology on a torus,  is observed to exhibit small
finite-size corrections and yields a high-precision critical point in two dimensions
(2D) as $h_c \! =\! 3.044\, 330(6)$, significantly improving over the existing
results and nearly excluding the central value of the previous result
$h_c \! =\! 3.044\, 38 (2)$. At criticality, the fractal dimensions of the  loops are estimated as $d_{\ell\downarrow} (1{\rm D})\! = \! 1.37(1) \!\approx\! 11/8$ and
$d_{\ell \downarrow} (2{\rm D}) \! = \! 1.75 (3)$,  consistent with those for the
classical  2D and 3D O(1) loop model, respectively. An interesting feature is that
in 1D, both the spin-down and -up loops display the critical behavior in the whole
disordered phase ($ 0 \! \leq \! h \! < \! h_c$), having a fractal dimension
$d_{\ell} \! = \! 1.750 (7)$ that is consistent with the hull dimension
$d_{\rm H} \! = \! 7/4$ for critical 2D percolation clusters. The current worm
algorithm can be applied to  simulate other quantum systems like hard-core boson
models with pairing interactions.
\end{abstract}

\pacs{00000}

\maketitle

\section{\label{sec:intro}introduction}
The quantum transverse-field Ising model (QTFI) is a textbook model in quantum many-body physics and plays an important role in quantum phase transition\citep{Subir2011} and quantum information science\citep{Dutta2015}. The one-dimensional (1D) QTFI can be solved exactly \citep{Pfeuty1970}, and it has been widely used to test theoretical or numerical methods\citep{Vidal2007, Blote2002} and to study novel quantities like entanglement entropy\citep{Kitaev2006, Levin2006} and quantum fidelity susceptibility\citep{Albuquerque2010}. In higher dimensions, analytical results are scarce, and one has to rely on numerical or approximate methods. Many  methods have been developed, including transfer matrix method\citep{Kaya2010}, series expansion\citep{Syljuasen2002, Albuquerque2010}, continuous-time Monte Carlo approach \citep{Rieger1999,Prokofev1998,Ikegami1998, Blote2002,Isakov2003, Iglovikov2013}, tensor renormalization group method\citep{Xie2012}, density matrix renormalization group\citep{White1992}, projected entangled-pair states\citep{Verstraete2004a}, and machine learning method\citep{Carleo2017,Carrasquilla2017} {\it etc}. Nevertheless, to obtain a high-precision critical point still remains to be a  challenging task. To our knowledge, the best estimates of the critical point for the 2D QTFI are $3.044\, 2(4)$ in Ref.~\onlinecite{Albuquerque2010} and $3.044\, 38(2)$ in Ref.~\onlinecite{Blote2002}, achieved by \textit{stochastic series expansion} (SSE) and continuous-time Wolff cluster methods respectively.

In this work, we apply a worm algorithm to simulate QTFI in 1D and 2D. It is shown
that as pointed out in Ref.~\onlinecite{Prokofev2001,Deng2007}, the worm algorithm  exhibits
efficiency  comparable to cluster schemes. A high-precision estimate of the
square-lattice critical point is obtained as $3.044\, 330(6)$, significantly
improving the existing results and nearly excluding the central value of $3.044\, 38(2)$
\cite{Blote2002}. In the path-integral representation for the current worm
algorithm, a configuration can be regarded as a set of non-intersecting loops
constructed by ``kinks" for pairwise interactions and spin-down (or -up)
imaginary-time segments. Rich geometric properties are observed for these loops.
In particular, it is found that in 1D, the loops over a wide parameter range
exhibit scaling laws that are in the universality class of the classical 2D
percolation. Deep theoretical understanding is desired. Further,  a variety of
physical quantities, including the magnetic and the fidelity susceptibilities, are
examined.  

The rest of the paper is organized as following. Section~\ref{sec:model} explains
the current path-integral representation for the QTFI and the formulation of
the worm algorithm. The numerical results are presented in Sec.~\ref{sec:numres}.
A brief summary is given in Sec.~\ref{sec:summary}.

\section{\label{sec:model} Worm algorithm}

\begin{figure}[t]
	\includegraphics[width=1.00\linewidth]{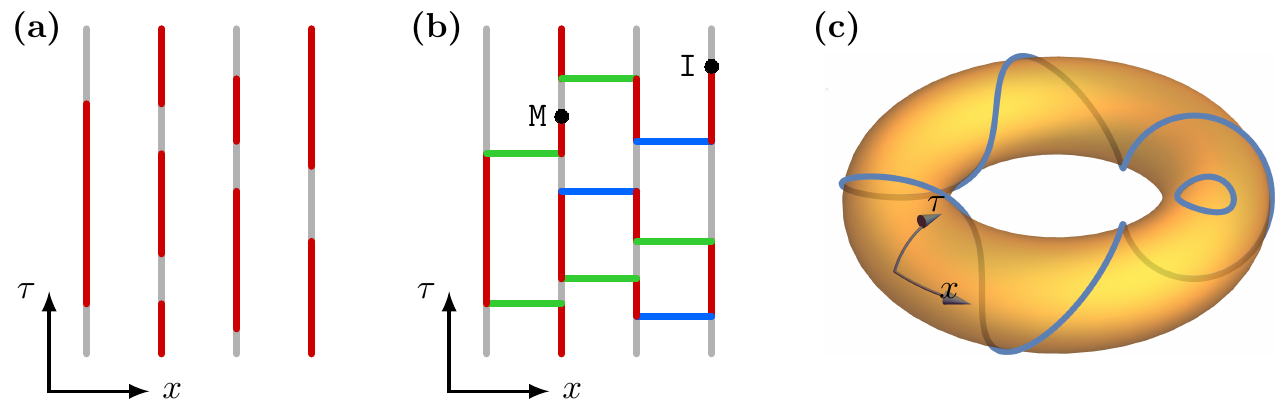}
	\caption{\label{fig:conf}(Color online)
	 Illustration of path-integral configurations. (a) for Eq.~\eqref{eq:hzzx}. Red
	 (gray) segments represent the spin-up (-down) state. (b) a $\mg$ configuration
	 for Eq.~\eqref{eq:greenfun} in the rotated basis, having an open path with the two ends marked as $\mi$
	 and $\mm$. Blue (green) lines are for pairing (hopping) kink. (c) a sketch of
	 $\mz$ configuration on the 1D torus, with a non-local loop of winding numbers
	 (${\cal W}_\tau=4, {\cal W}_x=1$). For simplicity, the lattice structure is not
	 shown.}
\end{figure}

The Hamiltonian of the QTFI on the $d$-dimensional cubic lattice is
\begin{equation}
	\label{eq:hzzx}
	\mathcal{H}=-t\sum_{\langle ij\rangle}\sigma_i^z\sigma_j^z-h\sum_i \sigma_i^x,
\end{equation}
where $\sigma_i^\alpha$ ($\alpha\!=\!x,\!z$) are Pauli matrices, $\langle ij\rangle$
represents nearest neighboring sites, $t \! >\!0$ is the ferromagnetic interaction
strength and $h$ is the transverse field. Taking the $\sigma^z $ spin component as
the expansion basis for the path-integral representation,  one can map a
$d$-dimensional QTFI onto a $(d \! +\!1)$-dimensional classical system, for
which each lattice site has a continuous line of spin segments; see
Fig.~\ref{fig:conf} (a) for an example. This continuous dimension is called the
imaginary-time ($\tau$) direction, along which the spin state $\sigma_i^z$ can be
flipped by the $\sigma_i^x$ operator but must satisfy the periodic condition. The
length of the $\tau$ dimension is the inverse  temperature $\beta = 1/k_{\rm B}T$
(the Boltzmann constant is set as $k_{\rm B} =1$ from now).

To formulate a worm algorithm that is effective for configurations of closed loops,
we choose the external-field direction as the expansion basis and rewrite
Hamiltonian~\eqref{eq:hzzx} as
\begin{equation}
	\label{eq:hxxz}
	\mathcal{H} \equiv K+U =-t\sum_{\langle ij\rangle}\sigma_i^x\sigma_j^x
	-h\sum_i \sigma_i^z   \; .
\end{equation}
As a result, $U$ and $K$
are respectively the diagonal and the non-diagonal terms. The pairwise interactions
$K=- t \sum_{\langle ij\rangle}\sigma_i^x\sigma_j^x$ can be further expressed in
terms of the raising and lowering spin operators,
$\sigma^{\pm} = (\sigma^x \pm i \sigma^y)/2$, as
\begin{eqnarray}
    \label{eq:eq3}
     K  &\equiv &  K_1+K_2 \\
         & = &-t\sum\limits_{\langle ij\rangle}(\sigma^+_i\sigma^-_j+{\rm H.c.})
				 -t\sum\limits_{\langle ij\rangle}(\sigma^+_i\sigma^+_j+{\rm H.c.}) \nonumber \; .
 \end{eqnarray}
The term $K_1$ flips a pair of opposite spins and thus the total magnetization is
conserved along the $\tau$ direction, while $K_2$ flips a pair of spins of the same
sign. We note that with the Holstein-Primakoff transformation,
$b_i(b^\dagger_i)=\sigma^-_i(\sigma^+_i)$ and thus
$n_i\equiv b_i^\dagger b_i =(\sigma^z_i+1)/2$, the QTFI can be mapped onto a hard-core
Bose-Hubbard (BH) model with Hamiltonian
\begin{equation}
\label{eq:BH}
\mathcal{H} = -t \sum_{\langle ij \rangle} (b^{\dagger}_i b^{}_j + {\rm H.c.}) -t' \sum_{\langle ij \rangle} (b^{\dagger}_i b^{\dagger}_j + {\rm H.c.})
- \mu \sum_i n_i  \; ,
\end{equation}
where $t'=t$, the particle number $n_i=0,1$, and the chemical potential
$\mu\!=\! 2h$. In the language of the hard-core BH model, $K_1$ accounts for the
hopping of a particle, and $K_2$, which simultaneously creates/deletes a pair of
particles, represents the pairing of two neighboring bosons. For convenience, we
shall refer to $K_1$ and $K_2$ as the hopping and the pairing term, respectively.

With Eq.~(\ref{eq:eq3}), the partition function of Hamiltonian~\eqref{eq:hxxz} can
be formulated in the Feynman's path-integral representation (also called the
world line representation) as
\begin{eqnarray}
	\label{eq:parfun}
		\mathcal{Z} &= & \text{Tr}\left[e^{-\beta \mathcal{H}}\right] %
		= \sum\limits_{\alpha_0 }\langle\alpha_0|e^{-\beta
		\mathcal{H}}|\alpha_0\rangle \\
		&= & 
		\lim_{\scriptsize{d\tau=\frac{\beta}{n}}\atop\scriptsize{n\rightarrow\infty}}\sum\limits_{\{\alpha_i\}}
		\langle\alpha_0|e^{-\mathcal{H}d\tau}|\alpha_{n-1}\rangle %
		\cdots \langle\alpha_1|e^{-\mathcal{H}d\tau}|\alpha_0\rangle \nonumber \\
		& = & \sum\limits_{\alpha_0}\sum_{{\cal N} =0}^{\infty} %
		\int_0^\beta  \! \int_{\tau_1}^\beta \! \cdots \! \int_{\tau_{{\cal N}-1}}^\beta \! \prod_{k=1}^{\cal N} \! d\tau_k \; F(t, t', h)  \nonumber 
\end{eqnarray}
with the integrand function  
\begin{equation}
\label{eq:weight}
F(t, t', h) \! =  {t}^{{\cal N}_{\rm h}} {t'}^{{\cal N}_{\rm p}} \exp\left(- \! \int_0^\beta \! U(\tau)d\tau\right) \; ,
\end{equation}
where ${\cal N}_{\rm h}$ and ${\cal N}_{\rm p}$ are, respectively, the number of
hopping and pairing kinks (${\cal N}={\cal N}_{\rm h}+{\cal N}_{\rm p}$),
$|\alpha_i\rangle=|\sigma_1^z,\sigma_2^z,\cdots,\sigma_N^z\rangle$ is an eigenstate
in the $\sigma^z$ basis ($N$ is the total number of lattice sites).
Moreover, Eq.~(\ref{eq:parfun}) can be
graphically viewed as the summation/integration over configurations in the
$(d+1)$-dimensional space-time $\{i,\tau\}$, of which the statistical weight is 
\begin{equation} 
\label{eq:WeightZ}
W_{\cal Z} (t,t',h) = \prod_{k=1}^{\cal N} \! d\tau_k \,  F(t, t', h) \; .
\end{equation}
In such a representation, each lattice site has a line of spin segments, and at
imaginary time $\tau_k $ ($k=1,2,\cdots,{\cal N}$), a pair of neighboring spins is
simultaneously flipped either by a hopping term $K_1$ or by a pairing term $K_2$.
We shall call them the hopping or the pairing kink, respectively. Starting from an
arbitrary space-time point $(i, \tau)$,  one would construct a closed loop by
following spin-up (-down) segments and kinks. Thus, a configuration
effectively consists of closed loops.

An important ingredient of the worm algorithm is then to extend the configuration
space ${\cal Z}$ for Eq.~(\ref{eq:parfun}) by including two defects. For the QTFI,
the extended configuration space ${\cal G}$ is for the
spin-spin correlation function of the Pauli matrix $\sigma^x$:
\begin{equation}
	\label{eq:greenfun}
	\mathcal{G}(\mathbf{x}_{\mi},\mathbf{x}_{\mm},\tau_{_\mi},\tau_{_\mm}) \! = \!  \textrm{Tr}\left[ T_\tau\left( \sigma_{_\mi}^x(\tau_{_\mi})\sigma_{_\mm}^x(\tau_{_\mm}) e^{-\beta \mathcal{H}} \right) \right],
\end{equation}
where $T_\tau$ is the $\tau$-ordering operator. In addition to closed loops,
a path-integral configuration in the ${\cal G}$ space contains an open path
with two ending points; see an example in Fig.~\ref{fig:conf}(b). We shall refer to
the ending points as ``Ira" ($\mi$) and ``Masha" ($\mm$), and denote their
coordinates in the space-time as $({\mathbf x}_{\mi}, \tau_{\mi})$ and
$({\mathbf x}_{\mm}, \tau_{\mm})$. The statistical weight of the ${\cal G}$
configuration can be written as 
\begin{equation} 
W_{\cal G} = \frac{d \tau_\mi d \tau_\mm}{ \omega_G} \;  \prod_{k=1}^{\cal N} \! d\tau_k \,  F(t, t', h) \; ,
\label{eq:WeightG}
\end{equation}
where $F$ is given by Eq.(\ref{eq:weight}) and $\omega_G$ is an arbitrary positive constant.
When $\mi$ coincides with $\mm$, the open path forms a closed loop, and the
$\cal{G}$ space is reduced to the $\cal{Z}$ space.

The full configuration space for the worm-type simulation corresponds to the
combination of the $\cal{G}$ and the $\cal{Z}$ spaces. For ergodicity, the
simulation must be able to change the kink number, the space-time location of any
kink as well as of defects ($\mi, \mm$), and to switch configurations back and forward between the
$\mathcal{Z}$ and $\mathcal{G}$ spaces. We adopt the following three updates:
(a) create/annihilate defects ($\mi,\mm$), (b) move imaginary-time of defect $\mm$,
and (c) insert/delete a kink. The first operation switches configurations between the ${\cal Z}$
and the ${\cal G}$ spaces by creating or annihilating a pair of defects $(\mi,\mm)$.
The second updates the $\tau_{\mm}$ value, and the third changes the
${\mathbf x}_{\mm}$ value by inserting or deleting a kink. Except ``create defects
$(\mi,\mm)$," all updates only apply in the ${\cal G}$ space, and each of
them is chosen with an {\it a priori} probability given before simulation.

\begin{figure}[t]
	\centering
	\includegraphics[width=0.90\linewidth]{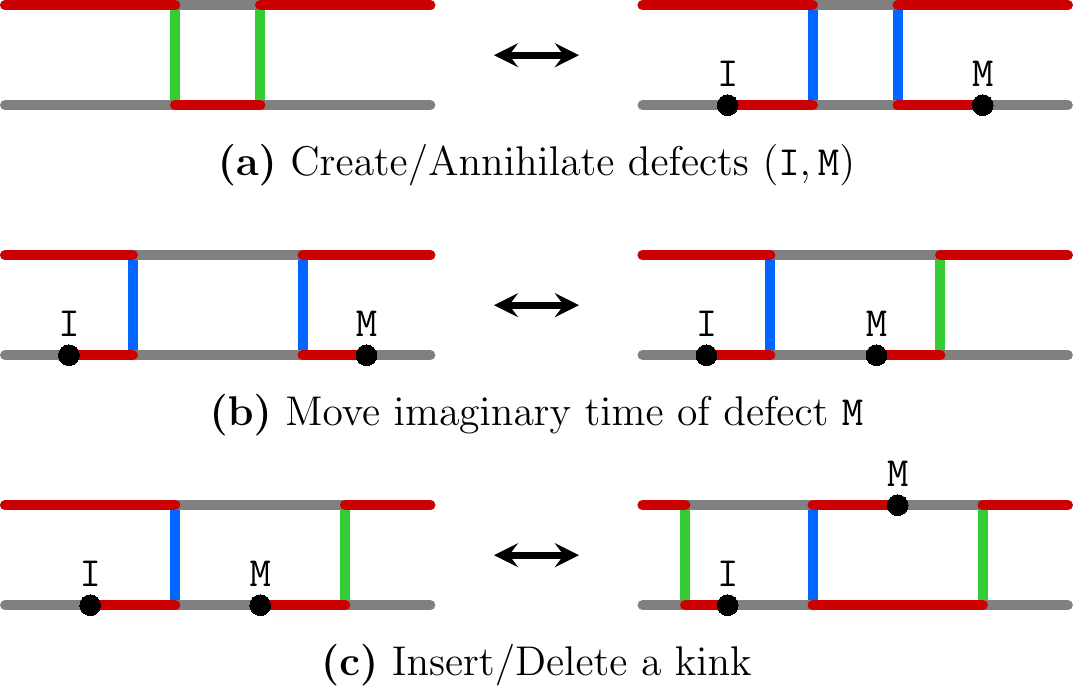}
	\caption{(Color online) The three updates. 
	}
	\label{fig:updatescheme}
\end{figure}

(a) {\it Create/annihilate defects} $(\mi,\mm)$. 
If the current configuration is in the $\mz$ space, ``Create defects $(\mi,\mm)$" is
the only possible update. 
     One randomly picks up a point $({\mathbf x}_\mi, \tau_\mi)$  from the whole space-time volume $\beta N \! = \! \beta L^d$,
     draws a uniformly distributed imaginary-time displacement $\delta \! \in \!  [-\tau_{\rm a}/2,\tau_{\rm a}/2)$ and $\delta\neq 0$ with the range $\tau_{\rm a} \sim O(1/h)$\cite{Note1}, 
     assigns ${\mathbf x}_\mm={\mathbf x}_\mi$ and $\tau_\mm  = {\rm mod} (\tau_\mi+\delta, \beta)$,
     and flips the spin state between defects $\mi$ and $\mm$.
     The $\beta$-periodicity is taken into account by the modular function.
     As illustrated in Fig.~\ref{fig:updatescheme}(a), the types (hopping or pairing) of kinks between 
      defects $\mi$ and $\mm$, if any, are interchanged during this operation.
     
	The update ``annihilate defects $(\mi,\mm)$," the reverse operation of ``create defects $(\mi,\mm)$," 
	is chosen with an {\it a priori} probability ${\cal A}_{\rm a}$ in the $\mg$ space. 
    It changes a $\mg$ configuration into a $\mz$ one by annihilating defects $(\mi,\mm)$ 
     and flipping the spin state inbetween. 
    This is possible only if $\mi$ and $\mm$ are on the same world line ${\mathbf x}_\mi={\mathbf x}_\mm$ 
	and their imaginary-time displacement $\min\left\{|\tau_\mi-\tau_\mm|, \beta-|\tau_\mi-\tau_\mm|\right\} \leq \tau_{\rm a}/2$.
	
	Accordingly, the detailed-balance condition reads as
	\begin{equation}
	\label{eq:db0}
	\frac{d \tau_\mi}{\beta N } \frac{d \tau_\mm}{\tau_{\rm a}} \cdot W_{\mz} \cdot  {\cal P}_{\rm crea} 
	= {\cal A}_{\rm a} \cdot  W_{\mg} \cdot {\cal P}_{\rm anni} \; , 
	\end{equation}
	where ${\cal P}_{\rm crea}$  (${\cal P}_{\rm anni}$) is the acceptance ratio for the ``create defects"  (``annihilate defects") operation,
	$W_{\mz}$ ($W_{\mg}$) is the statistical weight for the configuration before (after) the creation of defects, 
	and $d \tau_\mi/(\beta N)$ and $d \tau_\mm/\tau_{\rm a}$ account for the probability of choosing 
	the space-time location for $\mi$ and $\mm$, respectively.
	

	Making use of Eqs.~(\ref{eq:WeightZ}) and (\ref{eq:WeightG}), the acceptance probabilities for  the Metropolis filter can be calculated as	
	\begin{eqnarray}
	\label{eq:pacc:ca}
	P_{\rm crea} &=& \min\left[1, {\cal A}_{\rm a} \, \tau_{\rm a} \, \frac{\beta N }{ \omega_G} \frac{F_{\rm new}}{F_{\rm old} }\right] \\
	P_{\rm anni} &=& \min\left[1, \frac{1}{{\cal A}_{\rm a}}  \frac{1}{\tau_{\rm a}} \frac{\omega_G}{\beta N}  \frac{F_{\rm new}}{F_{\rm old} } \right] \; , \nonumber 
	\end{eqnarray}
	where  $F_{\rm new}$ and $F_{\rm old}$, given by Eq.~(\ref{eq:weight}), is respectively for the configuration
	after and before the corresponding operation. 
	Note that  the statistical-weight change, $F_{\rm new}/F_{\rm old}$, is mainly determined by the random displacement $|\delta| \leq \tau_{\rm a}/2$.
	As a result, the acceptance probabilities in Eq.~\eqref{eq:pacc:ca} can be optimized by tuning $\tau_{\rm a}$.
	

A natural choice for the relative weight is $\omega_G = \beta N$,
since the acceptance probabilities in Eq.~(\ref{eq:pacc:ca})  then hardly depend on $L$ and $\beta$. 
Physically, this is because the spin-spin correlation function $\mathcal{G}(\mathbf{x}_{_\mi},\mathbf{x}_{_\mm},\tau_{_\mi},\tau_{_\mm})$ has the space-time translation invariance so that the statistical weight of a ${\cal G}$ configuration should be normalized by the factor $1/\omega_G=1/(\beta N)$.
Further, with this choice, the number of Monte Carlo steps between two adjacent creations of ($\mi$, $\mm$), 
called the worm-return time, measures the ratio of the $\mathcal{G}$ space over the $\mathcal{Z}$ space,
and exactly gives the dynamic magnetic susceptibility of the QTFI which is stated explicitly in Sec.~\ref{subsec:wrt}.

 (b) {\it Move imaginary time of defect $\mm$.}
      The update, reverse to itself, is chosen with a probability ${\cal A}_{\rm b}$ in the $\mg$ space. 
      One randomly selects a time-displacement $\delta \in [-\tau_b/2, \tau_{\rm b}/2)$ and $\delta\neq 0$,
      assigns $\tau'_\mm = {\rm mod}(\tau_\mm+\delta,\beta)$ for the new temporal location of defect $\mm$,
      and flips the spin states inbetween; see Fig.~\ref{fig:updatescheme}(b). 
      The types of inbetween kinks are also  interchanged. 
	  The acceptance probability is 
	\begin{equation}
		P_{\rm move}=\min \left\{1, F_{\rm new}/F_{\rm old} \right\} \; .
	\end{equation}

 (c) {\it Insert/delete a kink.}
      Each operation is chosen with a probability ${\cal A}_{\rm c}$ in the $\mg$ space. 
      In ``insert a kink," one randomly chooses one of the $z_{d} \! = \! 2d $ neighboring world lines of ${\mathbf x}_\mm$, 
      say ${\mathbf x}'_\mm$, and updates the spatial location of  $\mm$ as $({\mathbf x}_\mm, \tau_\mm) \! \rightarrow \! ({\mathbf x}'_\mm, \tau_\mm)$.
      Meanwhile, one inserts a kink $k$ between world lines ${\mathbf x}_\mm$ and ${\mathbf x}'_\mm$
      at imaginary time $\tau_k = {\rm mod} (\tau_\mm+\delta,\beta)$, with a random displacement $\delta \in [-\tau_{\rm c}/2, \tau_{\rm c}/2)$ and $\delta\neq 0$. 
	  Further, the spin states between $\tau_\mm$ and $\tau_k$, on both  ${\mathbf x}_\mm$ and ${\mathbf x}'_\mm$, are flipped which causes the types of inbetween
	  kinks, linking  ${\mathbf x}_\mm$ and ${\mathbf x}'_\mm$, to stay the same.
	  However, the types of inbetween kinks are interchanged  if they link some other world lines to ${\mathbf x}_\mm$ or ${\mathbf x}'_\mm$.
       An example is illustrated in Fig.~\ref{fig:updatescheme}(c).
 
      In the reverse operation, ``delete a kink," one also picks up a random neighboring world line ${\mathbf x}'_\mm$ of $\mathbf{x}_\mm$ and
      moves $\mm$ as $({\mathbf x}_\mm, \tau_\mm) \rightarrow ({\mathbf x}'_\mm, \tau_\mm)$. 
      Further, one counts the number $n_{\rm k}$ of kinks that connect world lines ${\mathbf x}_\mm$ and ${\mathbf x}'_\mm$ 
      in the imaginary-time domain $[\tau_\mm-\tau_{\rm c}/2, \tau_\mm+\tau_{\rm c}/2)$. 
      If no kink exists $n_{\rm k} =0$, the operation is rejected.
     Otherwise, one randomly picks up one of the $n_{\rm k}$ kinks and deletes it, 
     and meanwhile flips the spin states on both world lines between $\tau_\mm$ and the imaginary time of the deleted kink. Besides types of kinks linking ${\mathbf x}_\mm$ or ${\mathbf x}'_\mm$ to other world lines are interchanged as well.

     The detailed balance condition of this pair of operations reads as
	\begin{equation}
	 {\cal A}_{\rm c} \cdot \frac{1}{z_d}  \cdot \frac{d \tau_k}{\tau_{\rm c} } \cdot W \cdot  {\cal P}_{\rm inse} 
	= {\cal A}_{\rm c} \cdot \frac{1}{z_d} \cdot  \frac{1}{n_{\rm k}} \cdot W_{+} \cdot {\cal P}_{\rm dele} \; ,  	
	\end{equation}
     where ${\cal P}_{\rm inse}$ (${\cal P}_{\rm dele}$) is for the acceptance ratio for ``insert a kink" (``delete a kink").
     The statistical weights $W$ and $W_{+}$, given by Eq.~(\ref{eq:WeightG}), are respectively for the configuration before and after inserting a kink. 
     The infinitesimal $d \tau_k$ on the left-hand side is cancelled by $W_{+}$, which has one more kink.    
     The acceptance probabilities are then 
     	\begin{eqnarray}
	\label{eq:pacc:id}
	P_{\rm inse} &=& \min\left[1,  \frac{\tau_{\rm c} }{ n_{\rm k}\!+\!1} \, \frac{F_{\rm new} }{F_{\rm old}}\right] \\
	P_{\rm dele} &=& \min\left[1,   \; \; \;\frac{ n_{\rm k}}{\tau_{\rm c} } \;\; \frac{F_{\rm new} } {F_{\rm old}} \right] \; , \nonumber 
	\end{eqnarray}
	where $n_{\rm k}$ denotes the number of inbetween kinks for the current configuration.  
	The denominator $n_{\rm k}+1$ in $P_{\rm inse}$ reflects an extra kink in the updated configuration.  

    The worm algorithm is then formulated as in Algorithm~\ref{algorithm},
    in which {\it a priori} probabilities satisfy ${\cal A}_{\rm a}+{\cal A}_{\rm b}+2{\cal A}_{\rm c}=1$. 
    It is mentioned again that the acceptance probabilities in the updates can be optimized by 
    tuning the ranges of random $\tau$-displacement,  $\tau_{\rm a}$, $\tau_{\rm b}$ and $\tau_{\rm c}$.
    As its analog for the classical Ising model which carries out a weighted random walk over the lattice, 
    the defect $\mm$ in this quantum Monte Carlo method effectively performs a random walk in the spacetime 
    and simultaneously updates the spin states it passes by.

For the conventional BH model which does not have the pairing term, 
the interchange between the hopping and pairing kink cannot be allowed. 
For ``create/annihilate defects" and ``move imaginary-time of $\mm$," the above illegal updates can be avoided when performing these operations {\it only} within a larger spin segment.
In  ``insert/delete a kink," the simplest remedy is that the proposed update is rejected as long as it leads to an illegal configuration,
giving a price that the acceptance probabilities are decreased by a factor of $O(1/h)$.
As a more sophisticated remedy, one can reformulate the operation in a way such that no illegal configuration would be introduced. 

Finally, for the computational efficiency, it is important to implement hash tables such that each operation 
is done within $O(1)$ CPU time.

\begin{figure}[t]
	\includegraphics[width=1.0\linewidth]{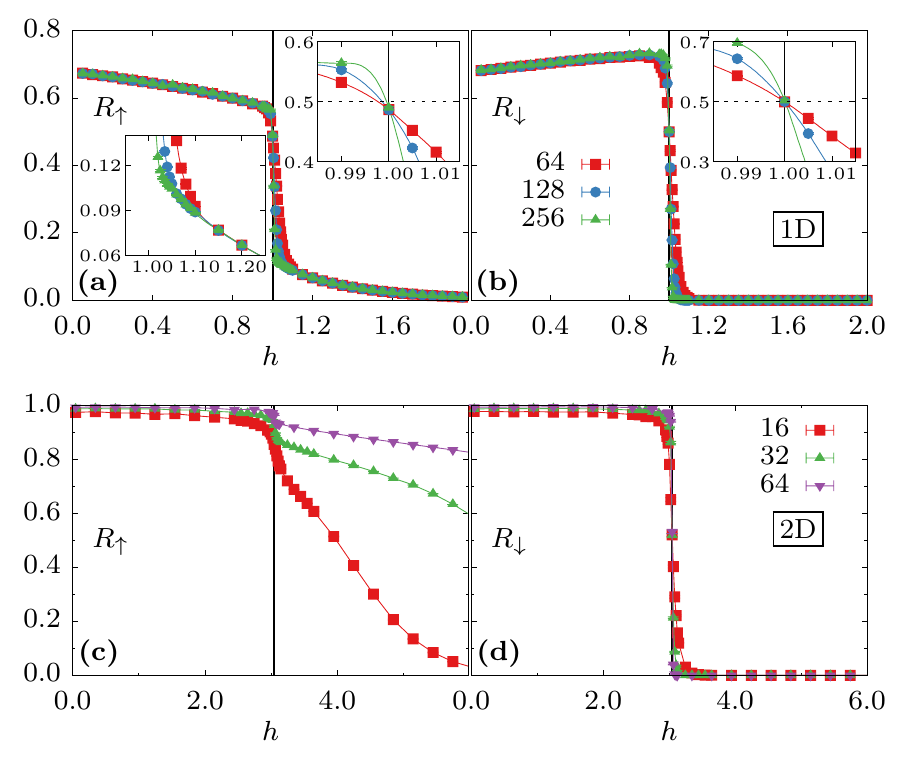}
	\caption{(Color online) Wrapping probabilities  $\mru$ and $\mrd$ versus the transverse field $h$. 
	The error bars are much smaller than the size of points. The vertical black lines indicate the critical point.
	(a), (b) are for 1D, and (c), (d) are for 2D. 
	 The inset plots of (a) and (b) show the curve near $h_c=1$.
	}
	\label{fig:wp12d}
\end{figure}

\section{\label{sec:numres}numerical results}

In the absence of the external field ($h=0$), the spin-up and -down states are fully balanced 
in the QTFI~(\ref{eq:hxxz}). As $h$ turns on,  
the system evolves into a disordered phase with the spin-down state being suppressed
and the spin-up order still not formed ($h<h_c$).
It enters into the spin-up ordered phase ($h>h_c$) through a second-order quantum phase transition.  
The critical point is exactly known as $h_c/t=1$ in 1D and numerically determined as 
$h_c/t \approx 3.044$ in 2D (square lattice).
Without loss of generality, the pairwise interaction is set as $t\!=\!1$ from now unless stated explicitly.

Using the worm algorithm, we simulate the 1D and 2D QTFIs with linear lattice size $L$ 
and inverse temperature $\beta=L$; the choice of $\beta=L$ is due to the dynamic critical exponent $z=1$ for the QTFI. 
Periodic boundary conditions are applied in each spatial direction, 
so that the lattice is essentially a torus.
The linear size is taken up to $L=512$ in 1D and $L=128$ in 2D, and
no severe critical slowing down is observed. 
A variety of geometric and physical quantities are sampled. 
To locate the phase transition $h_c$, we make use of the topological properties 
of the non-intersecting loops on the torus, 
instead of the scaling behaviors of physical quantities like the magnetic susceptibility. 

\subsection{Critical point}

Given a ${\cal Z}$ configuration, we record how many times ${\cal W}_i^{\ell} \geq 0 $ each loop $\ell$ winds along 
the $i$th direction $(i=1,2,\cdots,d)$, and calculate the total winding number ${\cal W}_i\! =\! \sum_{\ell} {\cal W}_i^{\ell}$ from all the loops;
see Fig.~\ref{fig:conf}(c) for an illustration.
A path-integral configuration is said to wrap along the $i$th direction as long as ${\cal W}_i > 0$.
This is indicated as ${\cal R}_i =1$;  otherwise, ${\cal R}_i =0$.
The average wrapping probability 
$R=(1/d) \sum_i \langle {\cal R}_i \rangle$ is then calculated, with $\langle \cdot \rangle $ representing the ensemble average. 
In the sub-percolating phase,  the loops are too small to percolate, and the $R$ value quickly drops to 0 as $L$ becomes larger. 
In the superpercolating phase, there is at least one giant loop with large ${\cal W}_i^{\ell}$, 
and the $R$ value rapidly converges to 1. 
At the percolation threshold, the $R$ values for different system sizes $L$ have an asymptotically common intersection with a non-trivial value 
between 0 and 1. 
In short, the wrapping probability $R$ is a dimensionless quantity characterizing the topological feature of loops on torus.
In many cases, such wrapping probabilities are found to exhibit small finite-size corrections, 
and  have been widely used for locating critical points~\cite{Newman2000,Martins2003,Feng2008,Wang2013,Hou2019}.

\vspace{2em}
\begin{algorithm}[H]
	\caption{Worm algorithm}
	\label{algorithm}
	\algsetup{indent=1em}
	\begin{algorithmic}
		\STATE {\bf BEGIN:} Given a $\mz$ configuration.
		\LOOP
			\IF {it is a  $\mz$ configuration}
				\STATE choose the ``create defects $(\mi,\mm)$" operation
			\ELSE
				\STATE  choose an operation with its {\it a priori}  probability 
				except ``create defects $(\mi,\mm)$"
			\ENDIF
			\STATE calculate the acceptance probability $P$ and carry out the operation with the probability $P$
		\ENDLOOP
	\end{algorithmic}
\end{algorithm}

\begin{figure}[t]
	\includegraphics[width=0.9\linewidth]{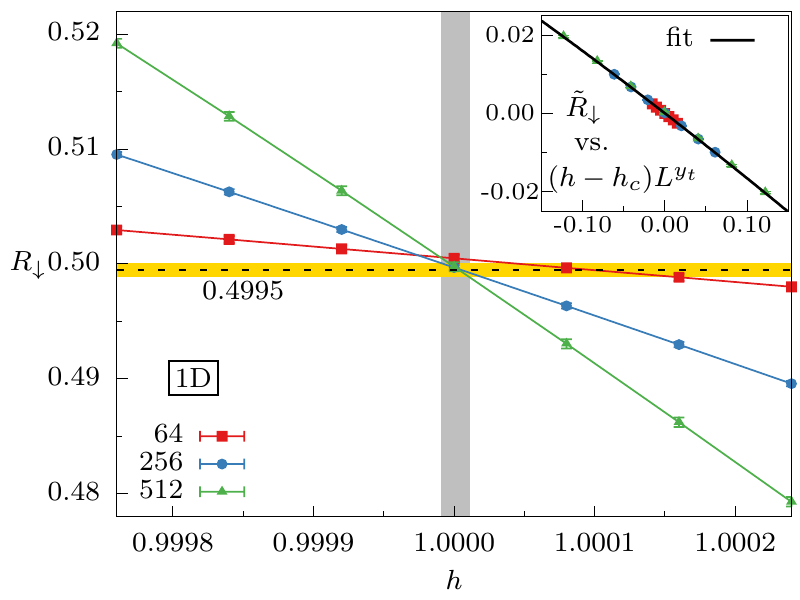}
	\caption{\label{fig:1dw}(Color online) Wrapping probability $R_{\downarrow}$ in 1D. 
	The gray band indicates an interval of $2\sigma$ above and below the estimate $h_c=1.000\, 001(5)$. 
	The yellow band indicates an  interval of $2\sigma$ above and below the estimate
	$R_c =0.4995(3)$.
	The inset shows  $\tilde{R}_{\downarrow}=R_{\downarrow}-R_c-b_i L^{y_i}$ versus  $(h-h_c)L^{y_t}$, 
	with $a_1=-0.1625(9), a_2=-0.34(13), b_i=4.3(1.5), y_i=-2$ and $y_t=1$.} 
\end{figure}

For the QTFI, two types of non-intersecting loops, spin-up ($\uparrow$) or -down ($\downarrow$) loops, can be constructed. 
The Monte Carlo (MC) results in Fig.~\ref{fig:wp12d} show that irrespective of the spatial dimension (1D or 2D), 
both the spin-up and -down loops display critical behaviors near the quantum critical point $h_c$.
In 1D, particularly rich behaviors are observed.
In the whole disordered phase ($0 \leq h <h_c$), 
both $R_{\uparrow}$ and $R_{\downarrow}$  have non-trivial values, $0< R_{\uparrow}(R_{\downarrow})<1$,
indicating the fractal structures of these loops.
At the transition point $h = h_c$, the $R_{\uparrow}$ and $R_{\downarrow}$ values have a sharp drop, 
which becomes infinitely sharp for $L \rightarrow \infty$.
For $h>h_c$, the $R_{\downarrow}$ value drops to 0, 
meaning that the spin-down loops are too small to percolate,
while the $R_{\uparrow}$ value converges to non-trivial value if $h$ is not too large, 
suggesting that  the spin-up loops still exhibit fractal properties.
Nevertheless, as $h$ is further increased,  the $R_{\uparrow}$ value also gradually approaches to 0.
This is understandable because
the number of kinks decreases when $h$ increases, and thus the spin loops  are less likely to percolate.
In 2D, the $R_{\uparrow}$ and $R_{\downarrow}$ values converge to 1 in the disordered phase $0 \leq h <h_c$,
suggesting a superpercolating phase both for the spin-up and -down loops.
For $h>h_c$, the $R_{\downarrow}$ value quickly reaches 0,  but $R_{\uparrow}$ seems to converge to 1 for $L \rightarrow \infty$.
At $h=h_c$, the $R_{\downarrow}$ value has a sharp drop,
and the derivative of $R_{\uparrow}$ with respect to $h$ probably also develops a singularity as $L$ increases.

Extensive simulations are then carried out at $h=h_c =1$ in 1D and $h=3.04435$ in 2D, 
and the data nearby are obtained by the standard reweighting technique\cite{Landau2014}.
The system size is taken as $L=16,64,256,512$ in 1D, with at least $4\times 10^8$ samples for each $L$,
and $L=8,16,32,64,128$ in 2D, with at least $1\times 10^8$ samples for each $L$.

\begin{figure}[t]
	\includegraphics[width=0.9\linewidth]{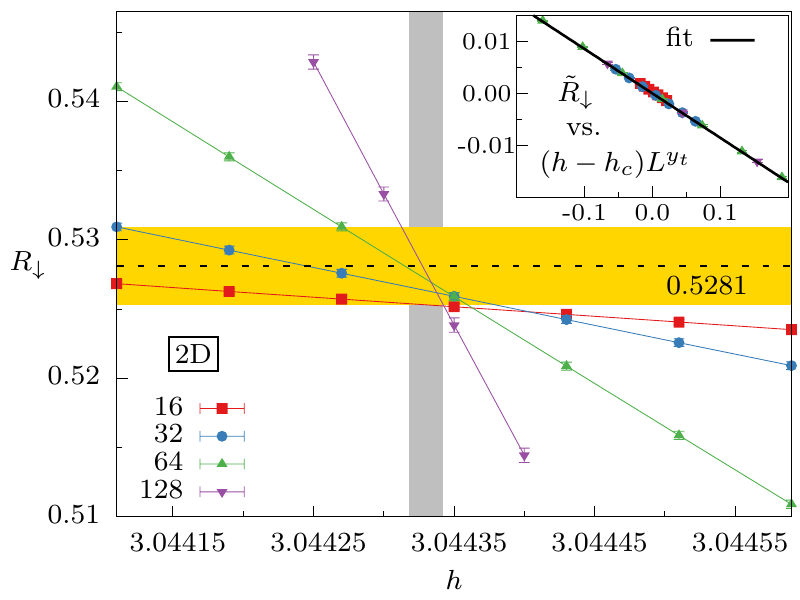}
	\caption{\label{fig:2dw}(Color online) Wrapping probability 
	$R_{\downarrow}$ in 2D. The gray band indicates an 
	interval of $2\sigma$ above and below the estimate $h_c=3.044\, 330(6)$. The yellow band indicates an interval of 
  $2\sigma$ above and below the estimation $R_c =0.528\, 1(14)$. The inset 
displays $\tilde{R}_{\downarrow}=R_{\downarrow}-R_c-b_iL^{y_i}$ versus
$(h-h_c)L^{y_t}$, with $a_1=-0.0855(8), b_i=-0.031(8), y_i=-0.821$ and $y_t=1.568$.} 	
\end{figure}

According to the least-squares criterion, the $R_{\downarrow}$ data, partly shown in Figs.~\ref{fig:1dw} and \ref{fig:2dw},
are fitted by 
\begin{equation}
	\begin{split}
		R_{\downarrow}=R_{\downarrow,c} & 
		+\sum_{k=1}^{2}a_k(h-h_c)^kL^{ky_t}+b_i L^{y_i}+b_2L^{y_2} \; .
	\end{split}
	\label{eq:Rfit}
\end{equation}
The thermal renormalization exponents are fixed at the known values in the classical ($d+1$) Ising universality--i.e., $y_t ({\rm 1D}) = 1$ 
and $y_t ({\rm 2D}) = 1.5868$~\cite{Blote2003}.
The term with $b_i$ comes from the leading irrelevant thermal field, 
which has the exponent $y_i ({\rm 1D}) = -2 $ and $y_i ({\rm 2D}) = -0.821$~\cite{Blote2003,Hou2019}.
The subleading correction exponents are set as $y_2 (\rm 1D) =-3$ and $y_2 (\rm 2D) =-2$. 
As a precaution, we gradually increase $L_{\rm min}$ and exclude the $L < L_{\rm min}$ data from the fit 
to see how the ratio of the residual $\chi^2$ to the degree of freedom changes with $L_{\rm min}$.

In 1D,  it is found that the MC data for $L_{\rm min}=64$ can be well described by Eq.~(\ref{eq:Rfit}) without the correction-to-scaling term ($b_2=0$).
The fit yields $h_c=1.000\, 001(5)$, in excellent agreement with the exact quantum critical point $h_c=1$.
Also, we have $R_{\downarrow,c}=0.4995(3)$, which suggests that it might exactly be $1/2$; see the inset of Fig.~\ref{fig:wp12d}(b).

\begin{figure}[!t]
	\includegraphics[width=0.9\linewidth]{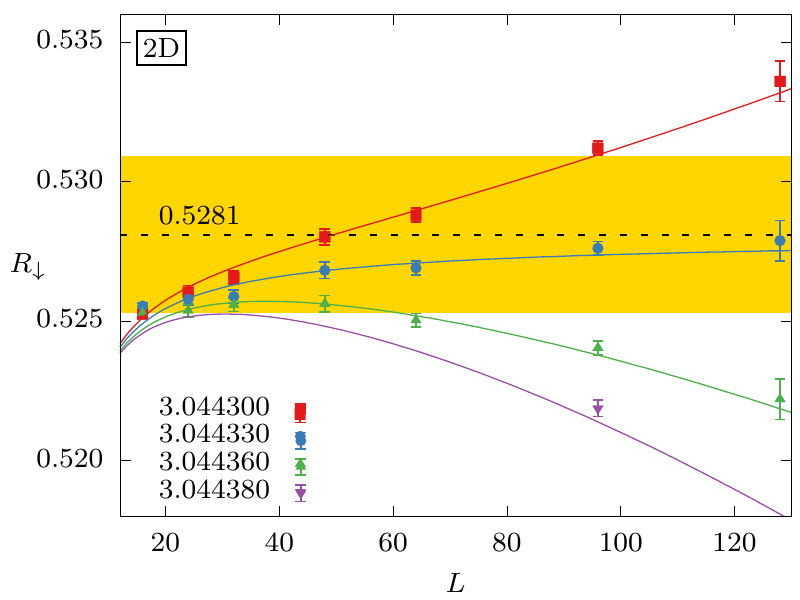}
	\caption{(Color online) Plot of $\mrd$ versus $L$ for various values of $h$ for
	the 2D QTFI. The yellow strip indicates an interval of $2\sigma$ above and below 
	the estimate $R_{\downarrow,c} =0.528\, 1(14)$. The solid lines are plotted according to the fitting result.} 
	\label{fig:directcp}
\end{figure}

In 2D, an eye-view fitting of the $R_{\downarrow}$ data in Fig.~\ref{fig:2dw} already gives
the critical point approximately as $h_c\approx 3.044\, 33$, with uncertainty at the fifth decimal place.
We find that it is sufficient to describe these data by Eq.~\eqref{eq:Rfit} with $a_2=0$ which means the fit is linear and for $L_{\rm min}=16$, $b_2$ can also be set to zero.
The fit gives $R_{\downarrow,c}=0.528\, 1(14)$ and $h_c=3.044\, 330(6)$.
To test the reliability of the value and the error bar of $h_c$, we plot in Fig.~\ref{fig:directcp} the $\mrd$ data 
against $L$  at some fixed $h$ near $h_c$. 
It can be seen that at $h\!=h_c = \!3.044\, 330$, the wrapping probability $R_{\downarrow}$ converges to a constant value 
within the $2\sigma$ shadow area in Fig.~\ref{fig:directcp}.
In contrast, as $L$ increases, the $R_{\downarrow}$ data for $h=3.044\, 300$ and  $3.044\, 360$  bend upward and downward,
respectively, suggesting that they are clearly away from the thermodynamic critical point. 
For $h=3.044\, 38$, which is the estimated central value of the critical point in Ref.~\onlinecite{Blote2002}, 
the downward bending is stronger, meaning that it cannot be the critical point. 
Table~\ref{tab:cp} gives a (incomplete) list of the existing results for $h_c$ in 2D.
It is clear that our estimate has the highest precision.

\begin{table}[!b]
	\caption{\label{tab:cp} Estimated critical point $h_c$  on the square lattice.}
\begin{threeparttable}
	\begin{ruledtabular}
		\begin{tabular}{ll}
			\addlinespace[2pt]
			Method &    $h_c$\\ [2pt]
			\hline \addlinespace[2pt]
			This work                              &   $3.044\, 330(6)$ \\ [2pt]
			CMC\citep{Blote2002}                   &   $3.044\, 38(2)$  \\ [2pt]
			SSE\citep{Albuquerque2010}             &   $3.044\, 2(4)$   \\ [2pt]
			S-W\citep{Rieger1999}                  &   $3.044(1)$       \\ [2pt]
			HOSVD(D=14)\citep{Xie2012}             &   $3.043\, 9$      \\ [2pt]
			iPEPS\citep{Orus2009}                  &   $3.04$           \\ [2pt]
			MERA\citep{Evenbly2009}                &   $3.075$          \\ [2pt]
			CTM\citep{Orus2012}                    &   $3.14$           \\ [2pt]
		\end{tabular}
	\end{ruledtabular}
	\begin{tablenotes}
	\footnotesize
\item[1] CMC: cluster Monte Carlo method
\item[] SSE: stochastic series expansion
\item[] S-W: Swendsen-Wang in continuous time
\item[] HOSVD: Tensor renormalization group method based on the higher-order singular value decomposition
\item[] iPEPS: infinity projected entangled-pair state
\item[] MERA: multiscale entanglement renormalization ansatz
\item[] CTM: corner transfer matrix
	\end{tablenotes}
\end{threeparttable}
\end{table}

We now briefly discuss the efficiency of the current worm algorithm, which is already reflected by 
the precision of the estimated critical point $h_c$.
For a quantitative evaluation, we calculate at criticality the integrated autocorrelation times $\tau_{_{\rm int}}$ for the energy ${\cal E}$, 
magnetization ${\cal M}$ and kink number ${\cal N}_k$, in the unit of MC sweeps. 
A MC sweep is defined such that on average, each imaginary-time spin line is updated by $\beta t$ times.
From the least-squares fitting $\tau \propto L^{z_{_\mathcal{O}}}$, we obtain the dynamical exponent as $z_{_{\cal E}}=0.38(3)$, $z_{_{\cal M}}=0.35(3)$ 
and $z_{_{{\cal N}_k}}=0.41(3)$ for 1D, and $z_{_{\cal E}}=0.28(3)$, $z_{_{\cal M}}=0.23(4)$ and $z_{_{{\cal N}_k}}=0.30(3)$ for 2D.
The efficiency of the worm algorithm 
is comparable to that of the Wolff-type cluster method~\cite{Blote2002}. 
Note that as the spatial dimension increases, all the values of $z_{_\mathcal{O}}$
decrease. 
From the numerical results of the worm algorithm for the classical Ising model~\cite{Prokofev2001,Eren2018}, 
we expect $z_{_\mathcal{O}}=0$ (without critical slowing down) for $d \geq d_c $, where $d_c=3$ is the upper critical dimensionality for the QTFI.

\begin{figure}[!t]
	\includegraphics[width=1.0\linewidth]{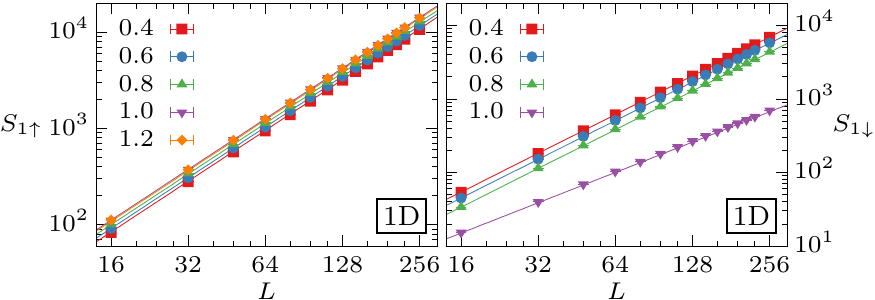}
	\caption{\label{fig:maxl}(Color online) 
	The  largest-loop size $S_{1\uparrow}$ and $S_{1\downarrow}$ versus $L$ at $h=0.4,0.6,0.8,1.0,1.2$ for 1D. 
	For $S_{1\uparrow}$, the straight lines have a slope $7/4$, irrespective of the $h$ value. 
	For $S_{1\downarrow}$, the lines have slope $7/4$ for $h<h_c$ and $11/8$ for $h=h_c$.}
\end{figure}

\subsection{Geometric properties of loops}
\label{subsec:geometric}
We have determined the quantum critical point $h_c$ with a high precision by locating the percolation threshold of the loop configurations.
Hereby, we shall further explore other geometric properties of the spin-up and -down loops at and away from $h_c$.
In the ${\cal Z}$ space, we measure the average length $S_1$ of the largest loop
and the probability distribution that a randomly chosen loop is of size $s$, i.e.,  $P(s,L) \equiv (1/N_\ell(L)) \partial N_\ell(s,L)/\partial s$, 
where $N_\ell(L) \sim \beta L^d $ is the total number of loops and 
$N_\ell(s,L)$ is the number of loops of size in range $(s,s+ds)$.
 
In 1D, one knows from Fig.~\ref{fig:wp12d} that for the whole region $0 \leq h \leq h_c$, 
both the spin-up and -down loops exhibit critical scaling behaviors. 
For the spin-up loops, such fractal properties further survive in the ordered phase $h>h_c$. 
In these cases, we expect that the largest-loop size scales as $S_1 \propto L^{d_{\ell}}$,
where $d_{\ell} < (1+1)$ is the loop fractal dimension, 
and that the loop-size distribution behaves as~\cite{Binder1976,MEFisher1967}
\begin{equation}\label{eq:distr}
P(s,L)\sim s^{-\tau} f(s/L^{d_\ell}),
\end{equation}
where $\tau$ is called the  Fisher exponent.
Moreover,  the exponents, $\tau$ and $d_\ell$, are related by the hyperscaling relation $\tau=1+(d+1)/d_\ell$.
The function $f(x\equiv s/L^{d_\ell})$ is universal and describes the finite-size cut-off of $s$ near $S_1 \sim L^{d_{\ell}}$.  

We simulate at $h=0.4,0.6,0.8,1.0,1.2$ and the results are shown in Fig.~\ref{fig:maxl}. 
The straight lines in the log-log plot suggest that indeed, the largest loop has a fractal structure. 
For the spin-up loops, irrespective of the $h$ value, the straight lines have the same slope approximately as $7/4$.
Further, the amplitude of the power law $S_{1\uparrow} \sim L^{7/4}$ increases as a
function of $h$, at least in the range of $ 0.4 \leq h \leq 1.2$.
In contrast, as $h$ increases, the largest-loop size $S_{1\downarrow}$ decreases and then drops to a significantly smaller value at $h=h_c$. 
Further, while the lines for $h<h_c$ still have a slope near $7/4$, the line for $h=h_c$ has a smaller slope which is about $11/8$. 
This suggests that the spin-down loops start with a dense and critical phase for $h<h_c$, 
experience a critical state at $h=h_c$, and then enter into a sparse phase containing enormous small loops. 
The $S_1$ data for both the spin-up and -down loops are fitted by 
\begin{equation}
	S_1 = L^{d_\ell}(a_0 + b_1 L^{y_1}) \; ,
	\label{eq:maxl}
\end{equation}
with different choices of the correction exponent $y_1=-0.5,-1.0$ or $-1.5$.
We find that the fits are rather stable, and the results are shown in Table~\ref{tab:1Ddl}.

\begin{figure}[!t]
	\includegraphics[width=0.9\linewidth]{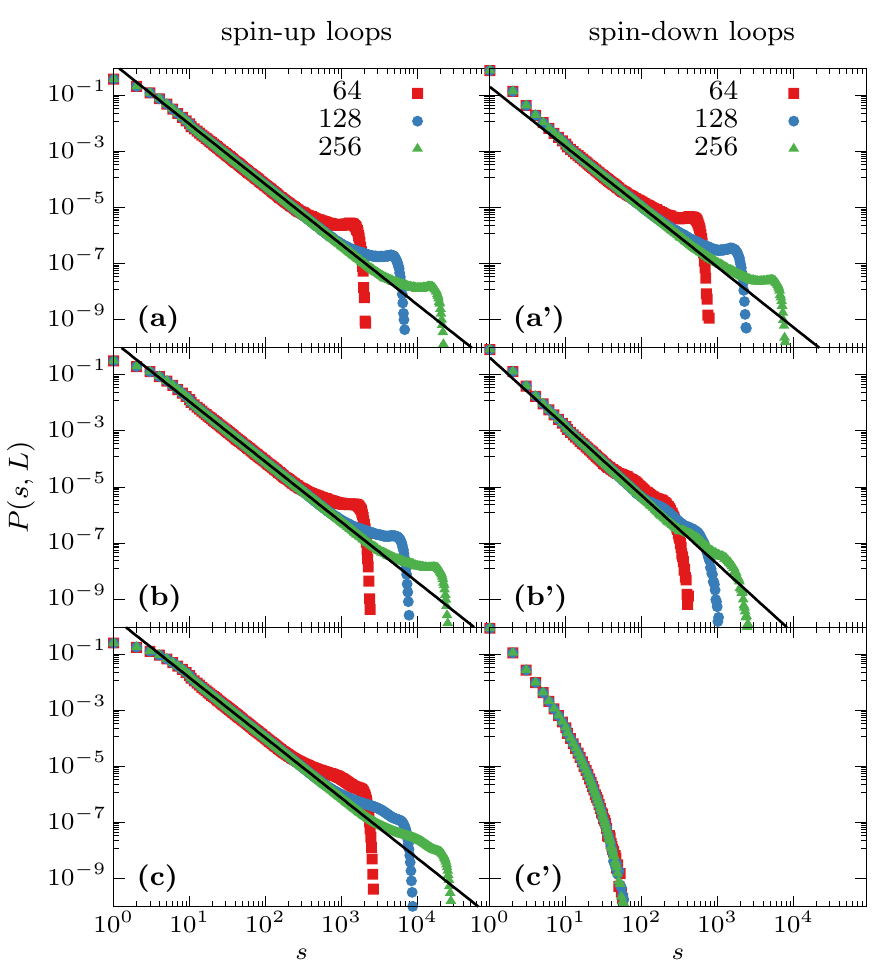}
	\caption{\label{1dhistogram}(Color online) 
	Loop-size distribution $P(s,L)$ for different $h$ values in 1D.
	The figures in the left (right) panel are for the spin-up (-down) loops, 
	and the first, second and third rows correspond to $h=0.8,1.0,1.2$, respectively. 
	The straight lines, with slope $-15/7$ or $-27/11$, are a guide for the eye.} 
\end {figure}

\begin{table}[!b]
  \caption{\label{tab:1Ddl}  Estimates of $d_{\ell\uparrow}$ and $d_{\ell\downarrow}$ at different $h$ in 1D.}
	\begin{ruledtabular}
		\begin{tabular}{cccccc}
			\addlinespace[2pt]
			$h$      &    0.4   &    0.6   &  0.8     &1.0($h_c$)&    1.2   \\ [2pt]
			\hline \addlinespace[2pt]
			$d_{\ell\uparrow}$  & 1.754(6) & 1.750(5) & 1.747(5) & 1.750(6) & 1.751(3) \\ [2pt]
			$d_{\ell\downarrow}$ & 1.751(5) & 1.749(7) & 1.750(7) & 1.37(1) &                   
		\end{tabular}
	\end{ruledtabular}
\end{table}

We notice that the configuration of the classical O($n$) loop model on the honeycomb lattice 
also consists of non-intersecting loops\cite{Liu2011,Saleur1987,Coniglio1989}. 
Moreover,  the O($n$) loop model with $n=1$ corresponds to the 2D Ising model, 
and has a hull/loop dimension as $d_{\rm hull}=11/8$ at the critical point $x_c=1/\sqrt{2+\sqrt{2-n}}=1/\sqrt{3}$ and $d_{\rm hull}=7/4$ in the dense phase $x>x_c$, 
where $x$ is the statistical weight for each loop unit\cite{Liu2011,Saleur1987,Coniglio1989}. 
These behaviors are very similar to those of the spin-down loops for the 1D QTFI.
Accordingly, we conjecture that in 1D, the fractal dimensions $d_{\ell\downarrow}(h=h_c)=1.37(1)$  
and $d_{\ell\downarrow} (h<h_c)=1.750(6)$ are exactly identical to $11/8$ and $7/4$, respectively.
We also conjecture that the fractal dimension $d_{\ell\uparrow}=1.747(5)$, which is independent of the $h$ value, is also exactly equivalent to $7/4$.
Further, it is noted that by the duality relation, the loops on the honeycomb lattice 
can be mapped onto the boundaries of the spin domains for the Ising model on the triangular lattice.
In the dense phase $x>x_c$, these domains are simply {\it critical} site-percolation clusters.
Therefore, we expect that the domains, enclosed by the spin-up or -down loops, 
are also fractal and have a fractal dimension
corresponding to that for critical Ising spin domains or percolated clusters in 2D.

\begin{figure}[!t]
	\includegraphics[width=0.9\linewidth]{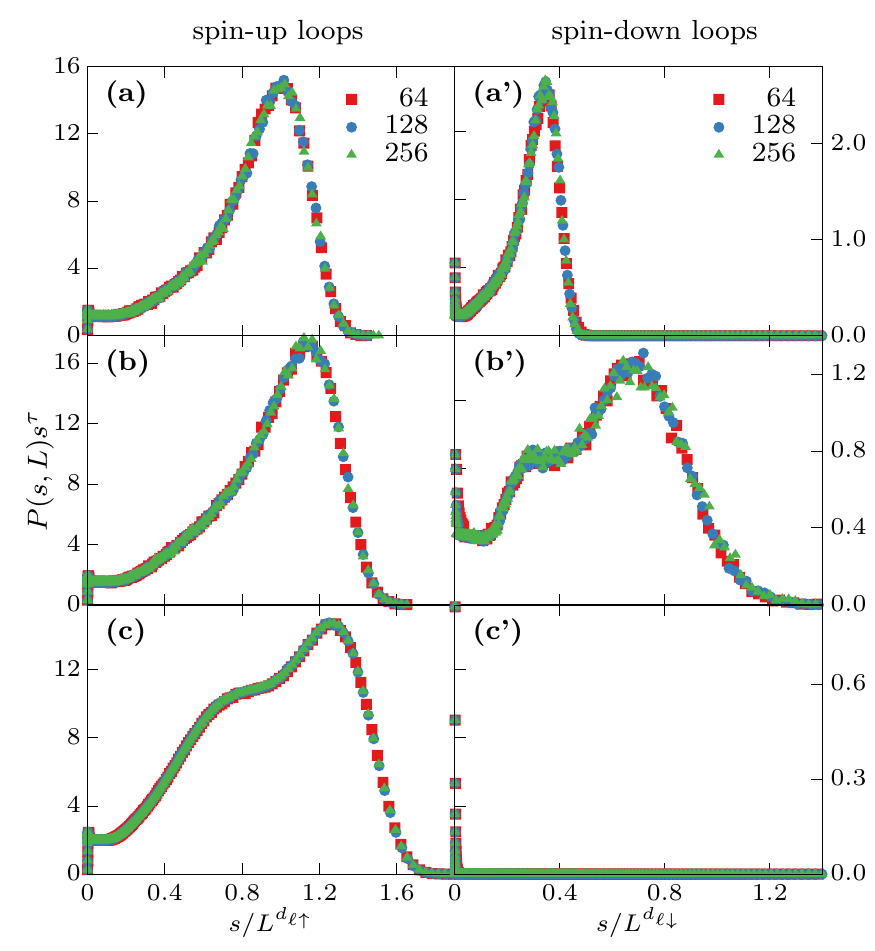}
	\caption{(Color online) $P(s,L)s^\tau$ versus $s/L^{d_\ell}$ in 1D.
	The plots in the left (right) panel are for the spin-up (-down) loops, 
	and the 1st, 2nd and 3rd rows correspond to $h=0.8,1.0,1.2$, respectively.
	The values of $d_{\ell\uparrow}$ and $d_{\ell\downarrow}$ are listed in Tab.~\ref{tab:1Ddl},
	and the $\tau$ value is calculated from the hyperscaling relation $\tau=1+(d+1)/d_\ell$.}
	\label{fig:1dcp}
\end{figure}

To further demonstrate the fractal structure of the spin-up and -down loops in 1D, 
we display in Fig.~\ref{1dhistogram} the MC data for the loop-size distribution $P(s,L)$. 
Indeed, one observes algebraically decaying behaviors, $s^{-\tau}$, for the spin-up loops with $h=0.8, 1.0, 1.2$
and for the spin-down loops with $h=0.8, 1.0$. 
The cut-off size of $s$ for the power-law scaling, due to finite-size effects, increases as the system size $L$. 
The hyperscaling relation $\tau = 1+(d+1)/d_\ell$ is well confirmed 
by the fact that the data for different $L$ collapse onto the straight lines with slope $-15/7$ or $-27/11$.
For the spin-down loops in the ordered phase $h=1.2$,  the $P(s,L)$ data for different $L$ drop quickly,
illustrating that the loop sizes are finite even in the thermodynamic limit $L \rightarrow \infty$.
We further plot $s^\tau P(s,L)$  versus $s/L^{d_\ell}$ in Fig.~\ref{fig:1dcp}.
With the values of $(d_\ell,\tau)$ as $(7/4,15/7)$ or $(11/8,27/11)$, 
the data for different $L$  collapse well onto a single curve, 
illustrating the universal feature of the cut-off function $f(x)$. 
It is interesting to see that for the spin-down loops at $h=h_c$, function $f(x)$  displays a two-peak structure (Fig.~\ref{fig:1dcp} (b')). 
We regard that the first peak at the smaller value of $x$ reflects the residual effect of 
the spin-down loops in the disordered phase $h<h_c$. 
Meanwhile, it is observed that for the spin-up loops with $h=1.2$, 
function $f(x)$  exhibits a shoulder feature on the smaller-$x$ side.
We expect that as $h$ increases, such a shoulder feature would become more pronounced 
and its location would move toward the value $x=0$. 
This is because that as $h$ is enhanced, 
the number of kinks will be gradually suppressed and the sizes of the spin-up loops will eventually start to decrease.
In the limiting case $h \rightarrow \infty$, all the spin-up loops will become individual imaginary-time lines with length $\beta$.

In 2D, the spin-down loops are fractal only at $h=h_c$, and the spin-up loops are always in a superpercolating phase.
The fit of the $S_{1\downarrow}$ data by Eq.~(\ref{eq:maxl}) gives $d_{\ell\downarrow}=1.75(3)$.
Again, this is in excellent agreement with the loop dimension $d_{\rm hull}=1.734(4)$ for the classical O($n=1$) loop model 
on the 3D  {\it hydrogen-peroxide} lattice~\cite{Liu2012}, on which the loops are also non-intersecting. 
As expected, for the spin-down loops at $h=h_c=3.044 \, 330$, the loop-size distribution $P(s,L)$ follows Eq.~\eqref{eq:distr}.

\begin{figure}[!t]
	\includegraphics[width=\linewidth]{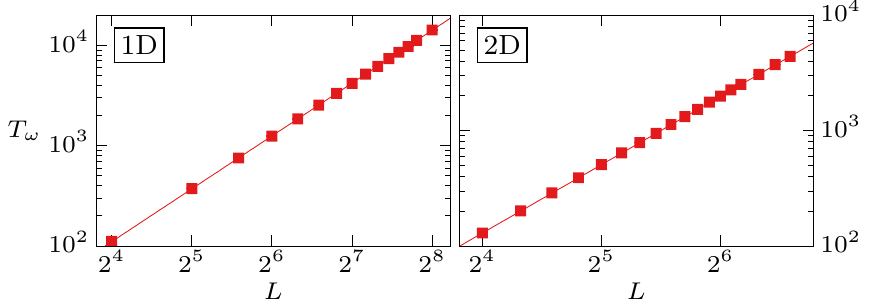}
	\caption{\label{fig:wrt}(Color online) Worm-return time at  criticality $h_c$. The straight lines, with slope $2y_h-(d+1)$, are a guide for the eye.}
\end{figure}

\subsection{Worm-return time}
\label{subsec:wrt}

The worm-return time $T_w$ is the average update steps between two adjacent
$\mathcal{Z}$ configurations in the markov chain MC simulation.
Mathematically, it can be expressed as the integral of spin-spin
correlation function~\eqref{eq:greenfun} over the lattice and the imaginary time as
\begin{eqnarray}
T_w	\! &=& \! \frac{1}{\omega_{G} \mathcal{Z}} \! {\rm\ Tr} \! \left[T_\tau \! \int_0^\beta \! \! \int_0^\beta \! {\rm d}\tau_{_\mi}{\rm d}\tau_{_\mm}  \! \sum_{\mathbf{x}_{_\mi},\mathbf{x}_{_\mm}} \sigma_{_\mi}^x(\tau_{_\mi})\sigma_{_\mm}^x(\tau_{_\mm}) 
		e^{-\beta\mathcal{H}}\right]  \nonumber \\
		\! &= &  \! \frac{1}{\omega_{G} \mathcal{Z}} \!  {\rm\ Tr} \! \left[\left(\int_0^\beta {\rm d}\tau \sum_{i} \sigma_{i}^x(\tau)\right)^2 e^{-\beta\mathcal{H}}\right] \nonumber  \\
		\! &= & \! \frac{1}{\omega_{G} \mathcal{Z}} \! {\rm\ Tr} \! \left[\left(\int_0^\beta {\rm d}\tau\ M^x(\tau)\right)^2 e^{-\beta\mathcal{H}}\right] \; .
		\label{eq:wrt}
\end{eqnarray}
With the choice of $\omega_G=\beta N$, the worm-return time $T_w$ is precisely equal to the dynamic magnetic
susceptibility $\chi^{xx}=\langle(\int_0^\beta {\rm d}\tau\ M^x(\tau))^2\rangle/\beta N$.
Figure~\ref{fig:wrt} shows the $T_w$ data at $h=h_c$ for both 1D and 2D, which are fitted by 
\begin{equation}
	T_w=L^{2y_h-(d+1)}(a_0+b_i L^{y_i}) \; .
	\label{eq:wrtfss}
\end{equation}
In 1D, the fit with $y_i=-2$ gives $y_h=1.876(2)$, in excellent agreement with the exact value $15/8$.
In 2D, we set $y_i=-0.821$~\cite{Blote2003} and obtain $y_h=2.484(4)$, which is again well consistent 
with the result $y_h=2.4816(1)$ for the classical 3D Ising model~\cite{Blote2003}.

\begin{figure}[!t]
	\centering
	\includegraphics[width=1.0\linewidth]{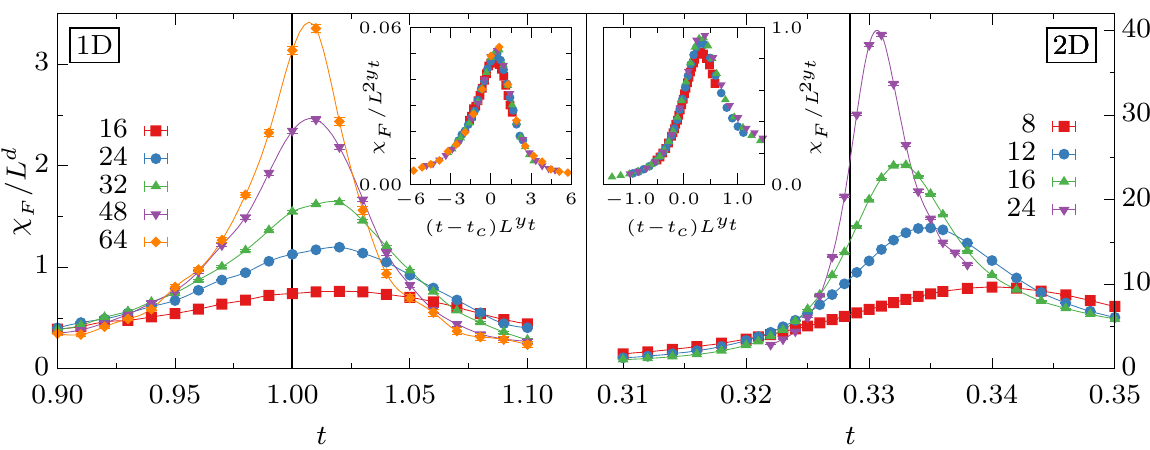}
	\caption{\label{chiF}(Color online) Fidelity susceptibility $\chi_{_F}(t)$ versus $t$, with (a) for 1D and (b) for 2D.
	The black solid lines indicate the critical points  $t_c(1{\rm d})=1$ and $t_c({\rm 2d})=1/3.044\, 330$. The insets show $\chi_{_F}/L^{2y_t}$ versus $(t-t_c)L^{y_t}$ which indicate the universality of the function $f_{\chi_{_F}}(x)$.}
\end{figure}

\subsection{Fidelity susceptibility}

It is well known that many systems can undergo quantum phase transitions without spontaneous symmetry breaking and thus
without a good definition of local order parameter. 
These phase transitions are beyond the Ginzburg-Landau paradigm,
and are difficult to be detected by conventional thermodynamic observables.
Fidelity susceptibility, a quantity proposed in the quantum information science~\cite{Nielsen2010},
has been shown to be useful for such a purpose~\cite{Zanardi2006,Albuquerque2010,Wang2015,Gu2009,Venuti2007}.
Consider a quantum phase transition driven by some given parameter $\lambda$ and let  $|\phi(\lambda)\rangle$ represent
the corresponding wave function, the fidelity $F(\lambda,\epsilon)$ of the system is defined as
the overlap between the wave functions with different values of $\lambda$--i.e.,  $F(\lambda,\epsilon)=|\langle\phi(\lambda)|\phi(\lambda+\epsilon)\rangle|$. 
Accordingly, the fidelity susceptibility $\chi_{_F}(\lambda)$ is calculated as:
\begin{equation}
	\chi_{_F}(\lambda) = -\left. \dfrac{\partial^2 \ln {\rm F(\lambda,\epsilon)}}{\partial {\epsilon}^2}\right|_{\epsilon=0} \; .
\end{equation}
For the QTFI, we hereby choose the driving parameter $\lambda$ to be the pairwise interaction $t$, 
which is conjugate to the number of kinks ${\cal N}_k$. 
Given a $\cal{Z}$ configuration, let ${\cal N}_{k,1}$ and ${\cal N}_{k,2}$ denote
the total number of kinks in the first-half imaginary-time domain $0 \leq \tau < \beta/2$ and the second-half one $\beta/2 \leq \tau < \beta$, respectively.
It can be shown by following Ref.~\onlinecite{Wang2015} that the fidelity susceptibility $\chi_{_F}(t)$ is proportional to the covariance of ${\cal N}_{k,1}$ and ${\cal N}_{k,2}$,
and can be written as 
\begin{equation}
	\chi_{_F}(t) = \dfrac{\langle {\cal N}_{k,1} {\cal N}_{k,2} \rangle-\langle {\cal N}_{k,1} \rangle\langle {\cal N}_{k,2}\rangle}{2t^2} \; ,
\end{equation}
where the external field $h$ is now set to be 1.
The MC data of $\chi_{_F}(t)$ for the 1D and 2D QTFIs are shown in Fig.~\ref{chiF}.
As expected, the $\chi_{_F}(t)$ data for each $L$ display a peak near the critical point $t_c$.
As system size $L$ increases, the peak location $t_{_L}$,  called the pseudo-critical point, 
moves toward the thermodynamic critical point $t_c$, and the peak itself becomes sharper with a smaller width. 
Following the standard finite-size scaling analysis, 
we expect that near the critical point $t_c$, the fidelity susceptibility $\chi_{_F}$ scales as
\begin{equation}
	\chi_{_F}(t,L) =L^{2 y_t}f_{\chi_{_F}}(L^{y_t}(t-t_c)) \; .
\end{equation}
Indeed, making use of $y_t({\rm 1d})=1$ and $y_t({\rm 2d})=1.5868$, we obtain a good collapse when plotting 
 $\chi_{_F}/L^{2y_t}$ versus $(t-t_c)L^{y_t}$, as shown in the insets of  Fig.~\ref{chiF}. 

For the fidelity  $F(\lambda,\epsilon)$, we can also choose the driving parameter to be the external field $h$ for the QTFI, 
which is conjugate to the $\sigma^z$-component magnetization ${\cal M}$.
In this case, we should consider the magnetization ${\cal M}_1$ for $0 \leq \tau < \beta/2$ and ${\cal M}_2$ for $\beta/2 \leq \tau < \beta$,
and the fidelity susceptibility $\chi_{_F}(h)$ would be proportional to the covariance of ${\cal M}_1$ and ${\cal M}_2$. 
At the critical point, we expect $\chi_{_F} \propto L^{2y_h}$.

While Fig.~\ref{chiF} illustrates the applicability of the fidelity susceptibility $\chi_{_F}$ as a tool for studying the quantum phase transition, 
it is worth mentioning that by calculating the covariance of two quantities of the same kind 
but in separated spatial/imaginary-time domains, $\chi_{_F}$ normally has large fluctuations.
Thus, to achieve a good statistics for $\chi_{_F}$ would  require extensive simulations.

\section{\label{sec:summary} Discussion}
We formulate a worm-type algorithm and study the QTFI in a path-integral representation 
in which configurations are sets of non-intersecting loops. 
By locating the percolation threshold of loop configurations via the so-called wrapping probability, 
we obtain a high-precision quantum critical point $h_c=3.044\, 330(6)$ for the QTFI on the square lattice.
These non-intersecting loops are further observed to exhibit rich geometric properties,
particularly in 1D, where both the spin-up and -down loops have fractal structures over a wide parameter range.
By examining the similarity of the scaling behaviors for the $d$-dimensional QTFI 
and for the $(d+1)$-dimensional classical O($n=1$) model, 
we conjecture that in 1D the two fractal dimensions are $d_{\ell \downarrow}(h_c)=11/8$ 
and $d_{\ell \downarrow}(h<h_c)=7/4$, and that in 2D,  $d_{\ell \downarrow}(h_c)=1.75(3)$ 
is equal to the hull dimension $d_{\rm hull}=1.734(4)$ for the classical 3D loop model.
The finite-size scalings of magnetic and fidelity susceptibilities are also examined.
It is confirmed that the fidelity susceptibility can be used to probe quantum phase transitions. 

Motivated by the fact that the classical O(1) loop model is a specific case of the O($n$) loop model with $n=1$, 
we can generalize the loop path-integral representation of the QTFI by giving each spin-down loop a statistical weight $n$.
As a consequence,  the partition function~(\ref{eq:parfun}) is generalized to be 
\begin{eqnarray}
	\mathcal{Z}(t,t',h,n) 
	= & & \sum\limits_{\{\alpha_0\}}\sum_{{\cal N} =0}^{\infty} %
	 \! \int_0^\beta \! \int_{\tau_1}^\beta \! \cdots \!  \int_{\tau_{{\cal N}-1}}^\beta \prod_{k=1}^{\cal N}  d\tau_i \nonumber \\
	& & n^{{\cal N}_{\ell \downarrow} } \, t^{\cal N_{\rm h}} \, {t'}^{\cal N_{\rm p}} \, \,  e^{-\int_0^\beta U(\tau) d \tau } \nonumber  \\
	= & & C\, \sum\limits_{\{\alpha_0\}}\sum_{{\cal N} =0}^{\infty} %
	 \! \int_0^\beta \! \int_{\tau_1}^\beta \! \cdots \!  \int_{\tau_{{\cal N}-1}}^\beta \prod_{k=1}^{\cal N}  d\tau_i \nonumber \\
	 & & n^{{\cal N}_{\ell \downarrow} } \, t^{\cal N_{\rm h}} \, {t'}^{\cal N_{\rm p}} \, (e^{-2h})^{\mathtt{S}_{\ell \downarrow}}\; ,
	\label{eq:parfungen}
\end{eqnarray} 
where $C=e^{h\beta N}$, ${\cal N}_{\ell \downarrow}$ specifies the number of spin-down loops 
and ${\cal S}_{\ell \downarrow}$ is the total length of spin-down loops.
We expect that for $t=t'$, the phase transition of such a ``quantum O($n$) loop" model in $d$ dimensions 
will belong to the same universality class as that for the classical O($n$) loop model in $(d+1)$ dimensions.
In 1D, we further expect that the exact value of the quantum critical point $h_c(n)$ can be obtained 
for the``quantum O($n$) loop" model, and that the spin-down loops would exhibit rich geometric properties 
both at criticality $h_c(n)$ and in the disordered phase $h<h_c$. 
In particular, for $(d=1,n=2)$, the phase transition would be of the celebrated Berezinskii-Kosterlitz-Thouless topological transition. 
All these expectations can be explored by the current worm-type algorithm, and remain to be a future work.

The efficiency of the current worm algorithm implies its broad applications 
in a variety of spin and hard-core systems. 
A straightforward application is to simulate the QTFI on other lattices regardless of dimensionality.
For the high dimension $d\geq 3$, one expects very minor or absent critical slowing down, 
and thus interesting logarithmic corrections can be examined.
It can be also of significant relevance in solid-state experiments, 
since the pairing terms $\sigma_i^+\sigma_j^+$ and  $\sigma_i^- \sigma_j^-$ are found to occur in frustrated quantum materials due to the dipolar-octupolar doublets\cite{Huang2014,Hatnean2015,Bertin2015,Lhotel2015,Xu2015,Anand2015,Benton2016,Xu2016,Anand2017,Dalmas2017,Sibille2015,Li2017,Gaudet2019,Gao2019,Li2019}.
Further, in addition to the external field $h$, one can introduce pairing interaction $\sigma_i^z\sigma_j^z$
along the $\sigma^z$ direction, which can be either ferromagnetic or anti-ferromagnetic. 
This allows the worm-type study of quantum spin systems with geometric frustration with respect to the $\sigma^z$  component.
In combination with the so-called clock Monte Carlo method~\cite{Manon2019}, one can even study spin systems with long-range $\sigma_i^z\sigma_j^z$ interaction 
without heavy computational overhead. 
Finally, we mention that a similar worm algorithm has recently been used in the SSE representation of 
the hard-core bosonic Hubbard model with pairing terms~\cite{Heng2019}.

\begin{acknowledgments}
	We acknowledge Nikolay Prokofiev for initializing the project and Xu-Ping Yao for technical help in preparing Fig.~\ref{fig:conf}(c).
	This work is supported by the National Natural Science Foundation of China under Grant No. 11625522.
\end{acknowledgments}


%

\end{document}